\documentclass[aps,prd,twocolumn,showpacs,superscriptaddress,footinbib]{revtex4-1}  
\usepackage{graphicx}  
\usepackage{dcolumn}   
\usepackage{bm}        
\usepackage{comment}
\usepackage{amssymb}   
\usepackage{amsmath}
\usepackage[usenames,dvipsnames]{color}
\usepackage[normalem]{ulem}

\newcommand{\pau}[1]{}

\begin{document}

\preprint{APS/123-QED}

  \title{Post-Newtonian Dynamics in Dense Star Clusters:
 \\Highly-Eccentric, Highly-Spinning, and Repeated Binary Black Hole Mergers}
  \author{Carl L.\ Rodriguez}
  \affiliation{MIT-Kavli Institute for Astrophysics and Space Research, 77 Massachusetts Avenue, 37-664H, Cambridge, MA 02139, USA}

       \author{Pau Amaro-Seoane}
          \affiliation{
Institute of Space Sciences (ICE, CSIC) \& Institut d'Estudis Espacials de Catalunya (IEEC)\\
at Campus UAB, Carrer de Can Magrans s/n 08193 Barcelona, Spain\\
Institute of Applied Mathematics, Academy of Mathematics and Systems Science, CAS, Beijing 100190, China\\
Kavli Institute for Astronomy and Astrophysics, Beijing 100871, China\\
Zentrum f{\"u}r Astronomie und Astrophysik, TU Berlin, Hardenbergstra{\ss}e 36, 10623 Berlin, Germany
            }

  \author{Sourav Chatterjee}
         \affiliation{Center for Interdisciplinary Exploration and Research in
    Astrophysics (CIERA) and Dept.~of Physics and Astronomy, Northwestern
      University, 2145 Sheridan Rd, Evanston, IL 60208, USA}

  \author{Frederic A.\ Rasio}
         \affiliation{Center for Interdisciplinary Exploration and Research in
    Astrophysics (CIERA) and Dept.~of Physics and Astronomy, Northwestern
      University, 2145 Sheridan Rd, Evanston, IL 60208, USA}

\date{\today}

\begin{abstract}

We present models of realistic globular clusters with post-Newtonian dynamics for black holes. By modeling the relativistic accelerations and gravitational-wave emission in isolated binaries and during three- and four-body  encounters, we find that nearly half of all binary black hole mergers occur inside the cluster, with about 10\% of those mergers entering the LIGO/Virgo band with eccentricities greater than 0.1. In-cluster mergers lead to the birth of a second generation of black holes with larger masses and high spins, which, depending on the black hole natal spins, can sometimes be retained in the cluster and merge again. As a result, globular clusters can produce merging binaries with detectable spins \emph{regardless} of the birth spins of black holes formed from massive stars.   These second-generation black holes would also populate any upper mass gap created by pair-instability supernovae.
%


\end{abstract}

\maketitle
%
%

\section{Introduction}
\label{sec:intro}

With the recent detections of five binary black hole (BBH) mergers, and one binary neutron star
merger, the era of gravitational wave (GW) astrophysics has arrived at last
%
%
%
\cite{Abbott2017c,Abbott2017,Abbott2017d,Abbott2016a,Abbott2016}.  Despite
significant theoretical work, the
origins of these systems, particularly the heavier BBHs,
remain an open question.  Both stellar evolution in isolated massive binaries
\cite[e.g.][]{Belczynski2010,Marchant2016,Podsiadlowski2003,Mandel2016a,DeMink2016}
and dynamical formation in dense star clusters \cite[e.g.][]{Sigurdsson1993,PortegiesZwart2000,OLeary2006,MillerLauburg09,Downing2010,Banerjee2010,Downing2011,Bae2014,Ziosi2014,Rodriguez2015a,Rodriguez2016a,Banerjee2017}
have been shown to produce merging BBHs similar to GW150914
\cite{Belczynski2016,Rodriguez2016b}. Understanding which formation pathways
are at play will be critical for the interpretation of GW data.  While many signatures of dynamical assembly
have been proposed, such as highly-eccentric mergers occurring in strong chaotic encounters \cite[][]{Samsing2014} or anti-alignment of the BH spins with the orbit
\cite[][]{Rodriguez2016c}, none of the BBH mergers detected so far by LIGO/Virgo have
displayed any of those signatures clearly \cite[see][]
{Amaro-SeoaneChen2016}.

What \emph{has} been displayed clearly in each BBH merger is the birth of a new
rapidly-spinning BH with a mass (almost) equal to the sum of its progenitor masses.  Many of these new BHs, particularly the remnants of GW150914,
GW170104, and GW170814, are significantly more massive than what is thought to form during
the collapse of a single star, where the pair-instability mechanism
limits the remnant BH mass to $\lesssim 50M_{\odot}$ \cite{Woosley2017}.
Were one of these mergers to occur in a dense star cluster, however, the merger
product could easily exchange into another BBH and merge again.  Because
of the distinct BH masses and spins in such second-generation (2G) mergers, it has
been suggested that such a population could be easily identifiable with future LIGO/Virgo detections
\cite{Fishbach2017,Gerosa2017}.  
%
%

In this letter, we present the first models of realistic globular clusters
(GCs) with fully post-Newtonian (pN) stellar dynamics. While relativistic $N$-body 
dynamics has been studied previously for highly idealized systems
\cite[e.g.,][]{Lee1993,Lee1992,Shapiro1985,Shapiro1985a,Rasio1989,Quinlan1987,Quinlan1989,Quinlan1990,KupiEtAl06,BremAmaro-SeoaneSpurzem2014} or open clusters \cite[e.g.,][]{Banerjee2017}, we show here for the first time using self-consistent
dynamical models of massive GCs that pN effects play a key role in assembling dynamically
the merging BBHs detectable by LIGO/Virgo. In our new pN models, we observe that
roughly half of all BBH mergers occur inside clusters, with a
significant fraction of those ($\sim 10\%$) merging with eccentricities greater
than 0.1 following GW captures. In-cluster mergers produce a second
generation of BHs that, if not ejected from the cluster through GW recoil, will dynamically
exchange into new binaries only to merge again. These 2G BBH mergers have
components with large spins and masses significantly beyond what is possible
from the collapse of a single star; they may be quite
common, with as many as $\sim$20\% of BBH mergers from our models having
components formed in a previous merger.  
%
%

Throughout this paper, we assume a
flat $\Lambda \rm{CDM}$ cosmology with $h= 0.679$ and $\Omega_M = 0.3065$
\cite{PlanckCollaboration2015}. 

\section{Post-Newtonian Dynamics}

\begin{figure}[tbh]
\centering
\includegraphics[scale=1, trim=0in 0in 0in 0.in, clip=true ]{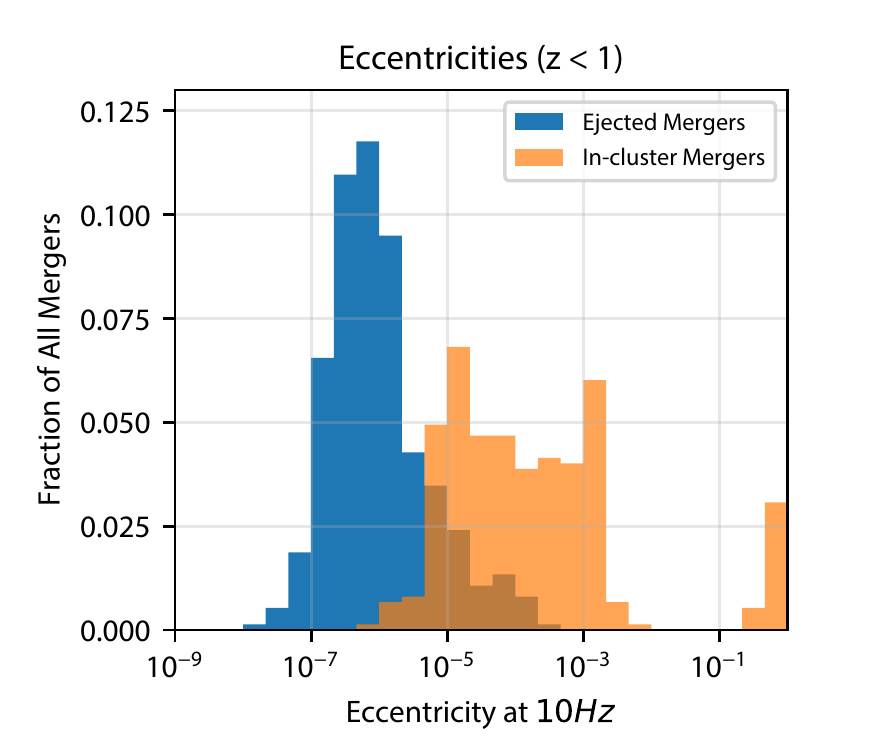}
\caption{The eccentricities of BBHs from the 24 GC models that merge at low redshifts.  We calculate the eccentricity when the BBH enters the LIGO/Virgo detection band at a (circular) GW frequency of 10Hz.   The distribution is clearly trimodal: the first peak corresponds to BBHs which merger after ejection from the cluster \cite[similar to][Figure 10]{Rodriguez2016a}.  The second peak corresponds to BBH mergers which occur in the cluster.  The final peak, at $e>0.1$, corresponds to in-cluster mergers which occur during a strong encounter, when the BBH enters the LIGO/Virgo band during a GW capture.  Note that the two distributions are normalized to the total number of mergers (in-cluster and ejected).}
\label{fig:ecc}
\end{figure}

We have computed the new GC models presented here using the
\texttt{Cluster Monte Carlo} (CMC) code.  CMC has been developed over many
years \cite{Joshi1999,Pattabiraman2013}, and includes all the
necessary physics for the long-term evolution of GCs,
including two-body relaxation \cite{Henon1971,Henon1971a}, single and binary
stellar evolution \cite{Hurley2000,Hurley2002,Chatterjee2010}, galactic tides, three-body binary formation
\cite{Morscher2012}, and three- and four-body gravitational
encounters via the \texttt{fewbody} package \cite{Fregeau2004,Fregeau2007}.
%
%
We have shown in \cite{Rodriguez2016} that CMC can reproduce with a high degree of fidelity both the global cluster
properties and BBH distributions computed with state-of-the-art direct $N$-body
simulations \cite{Wang2016}, while at the same time being at least two orders of magnitude faster (essential for the sort of extensive parameter-space study presented here).  Furthermore, CMC
has been upgraded \cite{Rodriguez2016a} to employ the most recent
prescriptions for stellar-wind-driven mass loss \cite{Vink2001,Belczynski2010a}
and compact-object formation \cite{Fryer2012}, allowing us to compare
our results directly to those of population synthesis studies for isolated binaries
\cite[e.g.][]{Belczynski2016}.

To incorporate pN effects into CMC, we make the following modifications. We
account for relativistic effects during three- and four-body encounters by adopting a modified version of the
\texttt{fewbody} code with pN accelerations up to and including the 2.5pN order.   
This code has been described in detail in 
\cite{Antognini2014,Amaro-SeoaneChen2016} and has been shown to conserve energy 
to 2pN order and to reproduce the inspiral times for compact binaries 
\cite{Peters1964}.  For BBHs which merge during an encounter, we perform a 
standard sticky-sphere merger, using detailed, spin-dependent fitting formulas 
from analytic and numerical relativity calculations  
\cite{Barausse2012,Kesden2008,Lousto2014,Berti2007,Campanelli2007,Tichy2008,Barausse2009,Kesden2008,Buonanno2008,Lousto2014,Tichy2008,Rezzolla2008,Gonzalez2007,Lousto2008,Lousto2012,Lousto2013}.  
{The new masses, spins, and recoil kicks are applied immediately during 
any merger, allowing us to model the retention of BHs by the cluster 
self-consistently.} See Supplemental Materials A for details, which includes 
references \cite{Wen2003,Blanchet2006,Sopuerta2007,Yunes2008,Gerosa2016,Jimenez-Forteza2017} . 
We initially assume all BHs from stellar collapse have no spins at birth 
($\chi_b = 0$, where $\chi$ is the dimensionless Kerr spin parameter), though we 
relax this assumption in Section \ref{sec:spin}.  For BBHs which do not merge during a
\texttt{fewbody} encounter, we directly integrate the orbit-averaged Peters 
equations \cite{Peters1964} for the change in semi-major axis and eccentricity 
due to GW emission.  This represents a departure from our previous work where we relied on the binary stellar evolution module \cite[BSE, ][]{Hurley2002}.  {By default, BSE only applies
GW energy loss to binaries with $a < 10R_{\odot}$.  This assumption
leads BSE to \emph{significantly} underestimate the number of GW-driven mergers
for binaries in a typical cluster, which can be highly eccentric and
very massive.  When accounting for GW energy
loss in eccentric binaries, the number of in-cluster mergers becomes comparable to the number
of merging binaries that are ejected from the cluster. This is a
significant improvement over previous results in the literature
\cite[e.g.,][]{Rodriguez2015a,Rodriguez2016a,Rodriguez2016b,Downing2010,Askar2016}, where
ejected BBHs dominated the merger rate in the local universe.  Semi-analytic approaches to cluster dynamics \cite[e.g.,][]{Antonini2016,Giesler2017} have reported significantly higher
fractions of in-cluster mergers, similar to those presented here, and have noted 
the possibility of multiple mergers in galactic nuclei \cite{Antonini2016}.}

We generate 24 GC models
covering a range of masses, metallicities, galactocentric  distances,
and virial radii, similar to those observed in the Milky Way and beyond. These
initial conditions are identical to those from \cite{Rodriguez2016a}, allowing
us to explicitly compare our pN results to those in the literature.   Our physics for single and binary stellar
evolution is nearly identical to \cite{Rodriguez2016a}.  We have added a
prescription for stellar mass loss via pulsational-pair-instability supernovae and stellar destruction via pair-instability supernovae. This physics, powered by the rapid production of
electron-positron pairs in the stellar core \cite{Woosley2017}, places a
well-understood upper limit on the masses of BHs that can from from the
collapse of a single star.  We take the limit from \cite{Belczynski2016a} of
$\sim 45M_{\odot}$ which is reduced to $\sim 40M_{\odot}$ via neutrino emission.  See
Appendix \ref{app:gc} for details.
This limit is in tentative agreement with the BH mass distribution measured by LIGO/Virgo
\cite{Fishbach2017a}.  In our simulations, no BH can be born with a mass above $40
M_{\odot}$ unless the BH or its stellar progenitor has undergone a dynamical
merger or mass transfer.   Finally, unlike previous studies
\cite{Rodriguez2015a,Rodriguez2016a,Rodriguez2016b}, we have not weighted our
models according to the distribution of observed GCs.  We
will explore more realistic sets of models in future work focusing specifically on LIGO/Virgo detection rates. In practice a more realistic weighting should make little difference, 
as our previously adopted weighting
scheme primarily selected BBHs from the most massive clusters, which also
contribute the majority of sources in our current grid.

\begin{figure*}[tb]
\centering
\includegraphics[scale=1, trim=0in 0in 0in 0in, clip=true ]{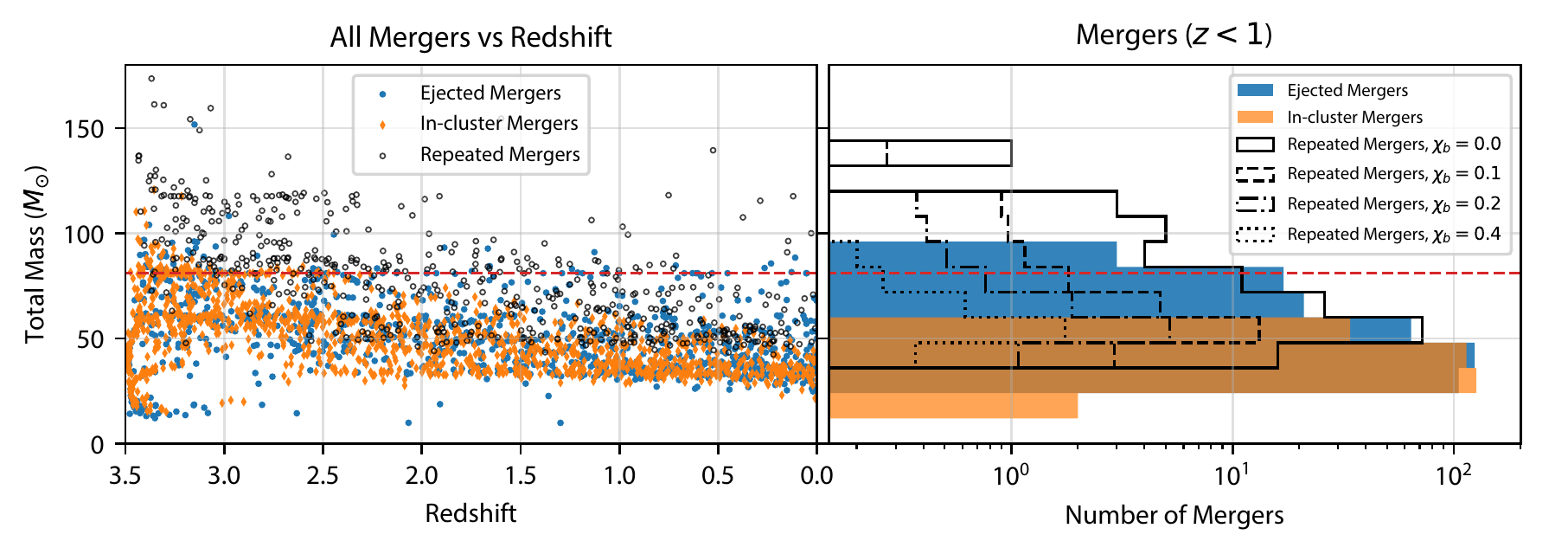}
\caption{The total mass of merging BBHs from all 24 GC models.  On the \textbf{left}, we show all mergers as a function of redshift, with the orange diamonds and blue points showing in-cluster and ejected mergers, respectively.  The black circles show 2G mergers (both in-cluster and ejected) which have at least one component that was formed from a previous BBH merger.  The \textbf{right} panel shows the mass distribution of these mergers at low redshifts ($z < 1$).    As the spins are increased from $\chi_b = 0$ to $\chi_b = 0.4$, the number of 2G mergers decreases significantly, as their progenitors were less likely to be retained in the cluster.  See the discussion in Section \ref{sec:spin}.  The red-dashed line indicates the maximum mass of first-generation BBHs ($\sim 81M_{\odot}$) with our assumed pair-instability supernova limit.  The handful of first-generation BBHs which merge above this are the result of either stable mass transfer or stellar collisions prior to BH formation.}
\label{fig:mz}
\end{figure*}

\section{In-Cluster Mergers}
\label{sec:ecc}

With the addition of the pN physics, we see a significant increase in the
number of in-cluster mergers.  Whereas before the number of in-cluster mergers
was a minor correction to the BBH mergers in the local universe \cite[0.06\% of
mergers at $z < 1$, see][]{Rodriguez2016a}, we now find that \emph{nearly
half} of mergers now occur inside the cluster.  For the 24 models considered here we
find a total of 2819 mergers, 55\% of which occur in the cluster.  At low redshifts ($z<1$), this
number decreases to 45\%, as the primordial
binaries which merged at early times after a common-envelope phase have merged many Gyr ago.
Compared to similar models without pN physics \cite{Rodriguez2016a},
the number of ejected BBH mergers at $z<1$ decreases by $\sim 20\%$ (496 versus
410).  However, the number of in-cluster mergers has jumped
significantly, from one to 338.  This increases the total number of mergers (in-cluster and ejected) by $\sim 50\%$.

{This increase in the number of BBH mergers occurring in the cluster 
primarily arises from properly accounting for GW emission for binaries 
regardless of their semi-major axis.  For example, a typical 
$30M_{\odot}+30M_{\odot}$ BBH is ejected from a GC with $a\sim0.4~\rm{AU}$ 
(roughly 10 times greater than the $a < 10R_{\odot}$ cutoff in BSE) after 
undergoing $\mathcal{O}(10)$ dynamical encounters \cite{Rodriguez2016b}.  During 
a typical encounter, the BBH semi-major axis will characteristically shrink 
while the orbital eccentricity randomly drawn from the thermal distribution, $p(e)de = 2e~de$  \cite{Heggie1975}.  These ``hardening'' encounters continue, shrinking the binary's semi-major axis until 
either the BBH is ejected from the cluster by the third body or until GWs 
drive the binary to merger.  The timescale for each BBH to merge can be roughly 
approximated by \cite[][]{Peters1964}:}

\begin{align}
	t_{\rm{GW}} \sim& \frac{e}{|de/dt|} \nonumber\\
	\sim& 400~\rm{Gyr} \left(\frac{a}{0.4\rm{AU}}\right)^4 \left(\frac{m_{\rm{BH}}}{30M_{\odot}}\right)^{-3} (1-e^2)^{7/2} \label{eqn:peterstimescale}
\end{align}

\noindent {As \eqref{eqn:peterstimescale} makes clear, a large eccentricity can significantly decrease the merger timescale.  For $e \gtrsim 0.95$ (roughly \%10 post-encounter binaries) $t_{\rm{GW}}$ will decrease by more than $10^3$, leading the BBH to promptly merge in the cluster.  On the other hand, for BBHs that never reach a high eccentricity, these encounters will continue to harden the binary until it is ejected from the cluster (where its eccentricity at ejection is set by a single draw from the thermal distribution).  Because the $(1-e^2)^{7/2}$ dependence in \eqref{eqn:peterstimescale} preferentially selects in-cluster mergers from a super-thermal distribution, we expect these mergers to have larger eccentricities than their ejected counterparts by the time they reach the LIGO/Virgo band. }

In Figure \ref{fig:ecc}, we show the eccentricity distribution of merging
binaries as they enter the LIGO/Virgo band (which we define as a circular
GW frequency of 10Hz).  We see {the expected} separation in
eccentricity between BBHs which merge in the cluster and those that merge after being ejected from the cluster.  For the in-cluster mergers, we also find a clear bimodality, with the
lower peak corresponding to isolated binaries that merge after a dynamical
encounter and the higher peak ($e > 0.1$) corresponding to sources which merge
\emph{during} the encounter via GW capture.  Although
previous work \cite{Samsing2014,Amaro-SeoaneChen2016,Samsing2017} has shown
through scattering experiments that such mergers are to be expected at the 1\%
level, this is the first work to show that these mergers occur in realistic GC
environments.  From our combined 24 models, we find that about 10\% of the in-cluster mergers
($\sim 3\%$ of all mergers) at $z< 1$ occur during these GW captures,
in good agreement with analytic work \cite{Samsing2017a}.

\begin{figure}[bth]
\centering
\includegraphics[scale=1, trim=0in 0.0in 0in 0.in, clip=true ]{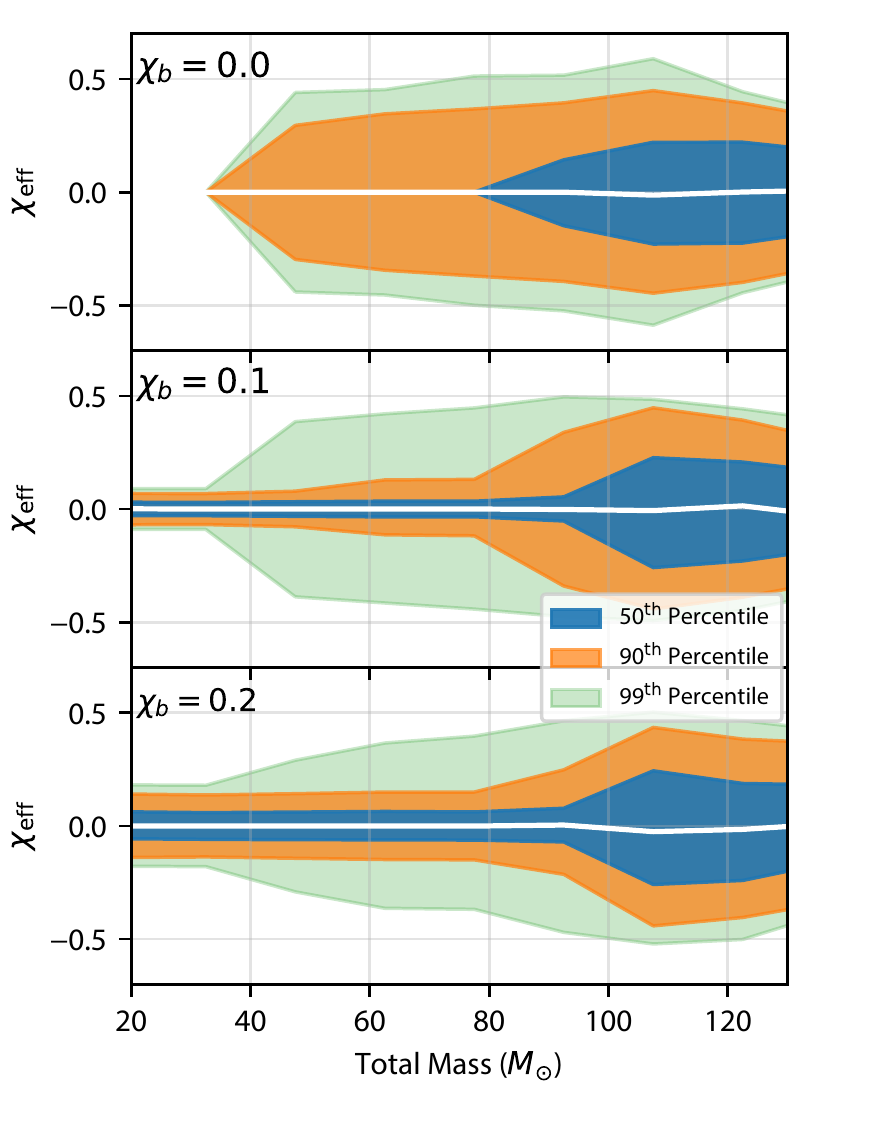}
\caption{The distributions of $\chi_{\rm{eff}}$ from BBHs that merge at $z<1$, 
divided into bins of $15 M_{\odot}$.  Each bin shows the median (white line), 
$50^{\rm{th}}$, $90^{\rm{th}}$, and $99^{\rm{th}}$ percentiles of 
$\chi_{\rm{eff}}$ for all BBH mergers with that mass. For each binary, we 
average over $N=10^3$ random spin orientations.  For the 2G mergers, we use 
$N=10^3$ times the probability of each component having been retained in the 
cluster following its earlier mergers (see discussion in Section 
\ref{sec:spin}).  As the birth spins ($\chi_b$) of the BHs are increased, 
the fraction of 2G BBHs retained in the cluster decreases; however, the overall 
magnitudes of $\chi_{\rm{eff}}$ increases, as the first generation of BBHs begin 
to produce mergers with measurable spins.  {Note that while the large 
spin magnitudes for BBHs with total masses above $80M_{\odot}$ does not depend on 
the birth spins, the number of mergers in that mass range decreases 
sharply with increasing $\chi_b$ (see Figure \ref{fig:mz}).}}
\label{fig:chieff}
\end{figure}

\section{Mergers over Cosmic Time}
\label{sec:mz}


In Figure \ref{fig:mz}, we show the mergers of BBHs as a function of
cosmological redshift.  What is immediately striking is that the mass distributions for in-cluster and ejected
binaries are significantly different at low redshifts.  This arises from
the delay times between formation and mergers for ejected BBHs.  When a BBH is
ejected from the cluster, it may still take several Gyr to merge in the field
\cite[see e.g.,][ and references therein]{BenacquistaDowning2013}.  Even for the
most massive clusters, the median inspiral time for ejected binaries is $\sim
10 \rm{Gyr}$ \cite[e.g.,][Figure 1]{Rodriguez2016a}.  In effect, the ejected BBHs which
merge today drew their components from the \emph{initial} distribution of BH
masses in the cluster, where the masses varied from $5M_{\odot}$ to
$40M_{\odot}$.  On the other hand, the in-cluster mergers have effectively no
delay time, and their components are drawn from the \emph{present-day}
distribution of BH masses in the cluster.  Because old GCs have ejected their
most-massive BHs many Gyrs ago \cite{Morscher2015}, the BBHs merging in the
cluster today are typically lower-mass than those that were ejected many Gyr ago.

Another interesting feature of
Figure \ref{fig:mz} is the presence of BBH mergers in the upper-mass gap, beyond the mass limit imposed by pair-instability supernovae.  The increased number of in-cluster mergers allows the GCs to produce
significant numbers of 2G BBH mergers, some of which will have
components above the maximum mass for BHs born from a single stellar collapse.  As these
systems can only be produced through multiple mergers, they will immediately be
identifiable as having arisen from a dynamical environment.  The rate of such mergers is small, but
LIGO/Virgo is more sensitive to mergers with more massive components \cite[the
detection horizon scales with the mass of the more massive component as
$m^{2.2}$,][]{Fishbach2017a}.  At the expected sensitivity for Advanced LIGO's
third observing run \cite{Abbott2016g}, a BBH with component masses of
$40M_{\odot} + 80M_{\odot}$ could be detected out to $z \sim 1$, encompassing a  comoving
volume of space three times larger than was observed during LIGO's second science run \cite{Chen2017}.

\section{Black Hole Spin and Recoil Kicks}
\label{sec:spin}

As a conservative assumption, we have assumed that all BHs in the cluster are
born with no intrinsic spin.  This is consistent with all but one
\cite[GW151226, ][]{Abbott2016} of the BBHs detected by LIGO/Virgo so far.  However, the presence of high BH spins, suggested by
observations of BH X-ray binaries \cite[see][for a review]{Miller2015}, can
radically change the results presented here: depending on the spin magnitudes
and orientations, merging BBHs can get kicks as high as 5000 km/s
\cite[e.g.,][]{Campanelli2007,Lousto2011,Lousto2012}, significantly larger than the escape speed of a
typical GC.  As a result, the 2G mergers shown in the left-hand panel of Figure
\ref{fig:mz} would not have formed if BHs are born with large
spins, since their components would not have been retained in the cluster \cite[][]{MillerLauburg09}.

We can estimate how the numbers in Figure \ref{fig:mz} would have changed under
different assumptions for BH birth spins.  For each repeated merger, we
calculate the probability that each of the components would have been retained
in the cluster given different birth spins.  This is done by computing the
recoil kicks over 1000 realizations of the spin orientations at merger.  The
probability of retaining each progenitor is simply the fraction of mergers for
which the recoil speed is smaller than the cluster escape speed where the
merger occurred.  For each 2G BBH merger, we take the product of the retention
probabilities for each component as the probability of that 2G merger
occurring.  We show the retention of these BBHs in the right panel of Figure
\ref{fig:mz} by weighting each 2G BBH merger by its retention probability.  As
expected, the number of 2G BBH mergers decreases as the birth spins of the BHs
are increased.  When $\chi_b = 0$, we find that $\sim$20\% of mergers at
$z<1$ are 2G mergers.  As the spins are increased, this number decreases, and
once $\chi_b = 0.4$, we observe $\mathcal{O}(1)$ 2G mergers, compared to the
672 first-generation mergers which occur at $z<1$.

These assumption have significant implications for the measurable spins of BBH
mergers. As shown by numerical relativity  \cite{Berti2008,Tichy2008,Lousto2010} and idealized
pN $N$-body simulations with spins
\cite{BremAmaro-SeoaneSpurzem2014}, repeated mergers of BBHs in clusters with
near-equal masses will tend to produce BHs with $\chi \sim 0.7$, 
{(assuming the initial spins are isotropically distributed 
\cite{Kesden2010})}.  But what
LIGO/Virgo is most sensitive to is not the spin magnitudes of the BBHs 
components \cite{Ajith2011,Purrer2016}, but the effective spin of the BBH, defined as the
mass-weighted projection of the two spins onto the orbital angular momentum:

\begin{equation}
\chi_{\rm{eff}} \equiv \left[\frac{m_1 \vec{\chi_1} + m_2 \vec{\chi_2}}{m_1 + m_2}\right] \cdot \hat{L},
\label{eqn:chieff}
\end{equation}

\noindent where $\hat{L}$ is the direction of the orbital angular momentum
and $\vec{\chi}_{1,2}$ are the
dimensionless-spin vectors for the BHs.  For dynamically-formed binaries, the
isotropic distribution of the orbit and spin vectors means that
\eqref{eqn:chieff} will be peaked at $\chi_{\rm{eff}} = 0$ with symmetric tails
whose extent depends on the BH spin magnitudes.  We show the distributions of $\chi_{\rm{eff}}$
in Figure \ref{fig:chieff}.  
When the initial BH spins are low, the 2G systems
are the only BBHs which merge with observably large spins.  The fraction of
systems with large spins increases as a function of total mass, since these
larger systems (particularly those beyond the pulsational-pair instability
limit) are predominantly formed through repeated mergers.  As the birth spins
are increased, the number of 2G mergers (with their characteristically large
spins) decreases as their components are more likely to be ejected from the
cluster during their first merger.  But the total number of BBH systems with
non-zero $\chi_{\rm{eff}}$ increases, as the first generation of BHs will now
form mergers with observable spins.  This result is key: one of the most
promising ways for identifying a dynamically-formed BBH merger is by the alignment of the spins, with
anti-aligned systems ($\chi_{\rm{eff}} < 0$) being a clear indicator of
dynamical formation \cite{Rodriguez2016c}.  These results indicate that
dynamical assembly in dense star clusters will inevitability produce a merger
with $\chi_{\rm{eff}} < 0$, regardless of the BH birth spins.

\section{Conclusion}
\label{sec:conc}

We have shown that the inclusion of pN effects can
have significant implications for BBH mergers from dense star clusters detectable by LIGO/Virgo.  
By accounting for GW emission
from isolated binaries and during three- and four-body dynamical encounters, we
find that a significant number of mergers occur in the cluster, and that
about 3\% of all mergers (and $\sim 10\%$ of in-cluster mergers) in our models will enter the LIGO/Virgo detection band
with high residual eccentricity ($e > 0.1$).  Because of this, GCs can potentially
produce a significant number of 2G BBH mergers with detectable spins and with masses 
larger than those produced through the collapse of single stars. Dynamics in dense star 
clusters can therefore produce BBH mergers with anti-aligned spins (a clear indicator of 
a dynamical origin)
\emph{regardless of the initial spins of first-generation BHs}: if natal BH spins
are large, then GCs can produce BBH mergers with $\chi_{\rm{eff}} < 0$ from 
first-generation systems.  If the spins are initially small \cite[as predicted by e.g.,][]{Amaro-SeoaneChen2016}, then the BBH merger products can often be retained in the cluster, forming a second generation of BBHs with large spins ($\chi \sim 0.7$).  
%
%

We thank Carl-Johan Haster, Michael Zevin, Johan Samsing, Davide Gerosa,
Salvatore Vitale, Chris Pankow, and Scott Hughes for useful discussions.  CR
%
%
is supported by a Pappalardo Postdoctoral Fellowship at MIT.  This work was
supported by NASA Grant NNX14AP92G and NSF Grant AST-1716762 at Northwestern
University.  PAS acknowledges support from the Ram{\'o}n y Cajal Programme of
the Ministry of Economy, Industry and Competitiveness of Spain, the
COST Action GWverse CA16104, and the CAS
President's International Fellowship Initiative.  CR thanks the Niels Bohr
Institute for its hospitality while part of this work was completed, and
the Kavli Foundation and the DNRF for supporting the 2017 Kavli
Summer Program.  CR and FR also acknowledge support from NSF Grant PHY-1607611 to the Aspen
Center for Physics, where this work was started.

\bibliographystyle{apsrev-carl}
\bibliography{Mendeley}

\appendix

\section{Post-Newtonian Physics}
\label{app:pn}

Our post-Newtonian (pN) scattering code was originally
presented in detail in \cite{Antognini2014,Amaro-SeoaneChen2016}, building upon the well-tested and commonly used
\texttt{fewbody} code \cite{Fregeau2004}.  \texttt{fewbody} computes 
strong binary-single and binary-binary encounters using an
$8^{th}$-order  Runge–Kutta Prince-Dormand integrator.  
Because the pN corrections to the
equations of motion can be expressed in terms of pair-wise position- and
velocity-dependent accelerations \cite[see e.g.,][]{Blanchet2006}, these terms
are easily incorporated into \texttt{fewbody}.   To incorporate both
conservative and dissipative relativistic effects (i.e., GW emission), the pN force terms 
are included in the code up to 3.5pN order,
though for the current study we only use the 1pN, 2pN, and 2.5pN terms.  This
code has been shown to conserve energy when dissipative effects are ignored,
and to reproduce accurately the inspiral times predicted by
\cite{Peters1964}.  See Appendices A1 and A2 in \cite{Antognini2014}.

We detect a merger of two BHs in \texttt{fewbody} using a standard sticky-sphere
collision criterion where the effective radii of the BHs are set to 5 Schwarzschild
radii ($10 G m / c^2$), {since the pN approximation begins to break down at separations of $r\sim10M$} \cite[e.g.,][]{Yunes2008}. Any two BHs touching within these radii are assumed to merge.  For
BBH mergers that occur during a \texttt{fewbody} integration, we determine
the {properties} of the merger remnant using the
 fitting formulas compiled in \cite{Gerosa2016}, based on extensive
 work to {parameterize the final mass (using the interpolation in 
 \cite{Barausse2012} based on results from \cite{Kesden2008,Lousto2014,Berti2007,Tichy2008}), final spin 
 (using the interpolation in \cite{Barausse2009} based on results from 
 \cite{Kesden2008,Buonanno2008,Lousto2014,Tichy2008,Rezzolla2008}, though see 
 \cite{Jimenez-Forteza2017} for more recent fits), and the recoil kick 
 (using the kick model from \cite{Campanelli2007} employing coefficients from 
 \cite{Gonzalez2007,Lousto2008,Lousto2012,Lousto2013}).}
Due to a typo in the original 
published version, we reproduce
the correct formula for the final mass here, while referring the reader to
\cite{Gerosa2016} for the prescriptions for the final spin and recoil kick. 
{We note that the \texttt{precession} software package described in 
\cite{Gerosa2016} does not contain this error}.
For the final mass, we use the interpolated fit between the test-particle
limits \cite{Kesden2008} and numerical relativity \cite{Lousto2014,Berti2007}
that was given in \cite{Barausse2009}.  This equation is incorrect in
\cite{Gerosa2016}, and should read

\begin{equation}
\frac{M_f}{M} = 1 - \eta (1 - 4\eta)(1-E_{\rm{ISCO}}) - 16 \eta^2 \left[p_0 + 4p_1 \tilde{\chi}_{\parallel}(\tilde{\chi}_{\parallel}+1) \right ]
\end{equation}

\noindent where $M_f / M$ is the ratio of the remnant mass to the binary mass, $\eta \equiv m_1 m_2 / (m_1+m_2)^2$, $p_0 = 0.04827$, $p_1 = 0.01707$ \cite[from][]{Barausse2009} 
and $E_{\rm{ISCO}}$ is the energy at the innermost-stable-circular orbit for a
Kerr BH:

\begin{align}
E_{\rm{ISCO}} &= \sqrt{1 - \frac{2}{3 r_{\rm{ISCO}}}}\\
r_{\rm{ISCO}} &= 3 + Z_2 - \rm{sign}(\tilde{\chi}_{\parallel}) \sqrt{(3-Z_1)(3+Z_1+2Z_2)}\\
Z_1 &\equiv 1 + (1-\tilde{\chi}_{\parallel}^2)^{1/3}\left[(1+\tilde{\chi}_{\parallel})^{1/3} + (1-\tilde{\chi}_{\parallel})^{1/3}\right]\\
Z_2 &\equiv \sqrt{3 \tilde{\chi}_{\parallel}^2 + Z_1^2}
\end{align}

\noindent and where $\tilde{\chi}_{\parallel}$ is:

\begin{equation}
\tilde{\chi}_{\parallel} \equiv \left[ \frac{m_1^2\vec{\chi}_1 + m_2^2 \vec{\chi}_2}{(m_1 + m_2)^2} \right]\cdot \hat{L}
\end{equation}

\noindent Note that $\tilde{\chi}_{\parallel}$ is not the same as 
$\chi_{\rm{eff}}$ in equation \eqref{eqn:chieff} {in the main text}.  We track the spin
magnitudes of any BHs {and update them accordingly when a merger occurs}.  
The spin directions during mergers (needed for the computation of the remnant mass, spin, and kick) are
assumed to be isotropically distributed on the sphere.  The recoil kick from
\cite{Campanelli2007} is decomposed into a component perpendicular to the
orbital angular momentum (to which both an asymmetric mass ratio and the spins
will contribute) and a component parallel to to the angular momentum, which
arises only from the spins.  We draw a random angle in the plane of the binary
for the perpendicular component and apply the two kicks in this coordinate
frame.  

During integrations of three- and four-body encounters, we do not stop the integration 
when two BHs merge, but instead continue until
\texttt{fewbody}'s standard termination criteria (which check for unbound
collections of objects and dynamical stability) are satisfied.  This allows us
to self-consistently track the outcome of mergers during dynamical encounters,
which can be important for the retention of BH merger products.

{
While we are applying the kicks self-consistently during the 
\texttt{fewbody} integration, we are not taking into account the residual 
eccentricity of the BBHs which merge during scattering encounters.  As 
eccentric, non-spinning mergers receive higher kicks than circular BBH mergers 
\cite[e.g.,][]{Sopuerta2007}, it is entirely possible that we are overestimating 
the retention of mergers which occur during three- and four-body encounters.  We 
do not implement these eccentric corrections here, since there exist no corresponding fitting 
formulae for eccentric BBH mergers with spinning components.}

For BH -- (non BH) star mergers, we do not consider any accretion of material from the
star, and assume that the entire star is tidally disrupted.  This is a
highly-conservative assumption, but is safest without a better treatment for
accretion and spin-up of compact objects (which, while potentially interesting, is
 beyond the scope of this study).

For isolated BBHs in the cluster, CMC had previously relied on BSE to determine
the change in eccentricity ($e$) and semi-major axis ($a$) at each timestep.
However, the evolution of $a$ and $e$, when averaged over the binary orbit,
depends strongly on the eccentricity \cite{Peters1964}:

\begin{align}
\left<\frac{da}{dt}\right> &= -\frac{64}{5} \frac{G^3 m_1 m_2 (m_1+m_2)}{c^5 a^3 (1-e^2)^{7/2}} \left( 1+\frac{73}{24}e^2 + \frac{37}{96}e^4 \right)\label{eqn:dadt}\\
\left<\frac{de}{dt}\right> &= -\frac{304}{15} e \frac{G^3 m_1 m_2 (m_1+m_2)}{c^5 a^4 (1-e^2)^{5/2}} \left( 1 + \frac{121}{304} e^2 \right)\label{eqn:dedt}
\end{align}

\noindent As BSE uses a forward-Euler integration for its equations, it can 
underestimate the merger time for binaries that reach very high eccentricities.
As an example, BSE will overestimate the merger time of a
$30M_{\odot}+30M_{\odot}$ binary with a semi-major axis of 0.1 AU and an
eccentricity of 0.9 by nearly 20\%. {As stated in the main text, the 
orbital eccentricity of a BBH is critical for the proper treatment of in-cluster 
mergers.}   
For isolated binaries in the cluster, we directly integrate equations
\eqref{eqn:dadt} and \eqref{eqn:dedt} using an $8^{th}$-order Runge-Kutta
Prince-Dormand integrator.  This is done outside of the BSE package, so that
any merging binaries use the same final mass, final spin, and recoil
fitting-formula that are employed for mergers in the \texttt{fewbody}
integrator.  As with mergers occurring during encounters, we terminate the
integration when the effective radii of the two BHs (5 Schwarzschild radii) touch. 
To determine the eccentricity for these systems in Figure
\ref{fig:ecc}, we integrate $de/da$, found by dividing equations
\eqref{eqn:dadt} and \eqref{eqn:dedt}.  
We integrate from the last-reported $a$ and $e$ in the cluster for each binary (which, for mergers in \texttt{fewbody}, is the $a$ and $e$ at the point of contact) to the semi-major axis corresponding to a circular GW
frequency $f_{\rm{GW}}=10\,$Hz (where $f_{\rm{GW}} = 2f_{\rm{orbital}}$).  Although
fitting formulae exist for the peak GW frequency of eccentric mergers \cite[e.g.,][]{Wen2003}, we do not use these
here, as several of these GW-capture BBHs form \emph{inside} the
LIGO/Virgo band, and therefore did not pass through the eccentric 10Hz threshold as an
isolated binary.

\section{GC Models}
\label{app:gc}

We evolve a 4x3x2 grid of GC models with different initial masses,
metallicities (assumed correlated with galactocentric distance), and virial radii.  
These clusters have
identical initial conditions to those considered and discussed in more detail 
in \cite{Rodriguez2016a}.  We
consider clusters with $2\times 10^5$, $5\times10^5$, $10^6$, and $2\times
10^6$ particles (single or binary stars) initially, covering a realistic range of GC 
masses. The initial positions and
velocities are distributed in phase space following a standard King
model \cite{King1966} with concentration $W_0 = 5$.  The virial radii of the
clusters are then scaled to $1$ or $2$ pc, representing a realistic range of initial sizes.  
The individual stellar masses are drawn from the range
$0.08\,M_{\odot}$ to $150\,M_{\odot}$ assuming a Kroupa initial-mass function (IMF)
\cite{Kroupa2001a}:

\begin{equation}
P(m)dm \propto m^{-\alpha}dm
\label{eqn:kroupa}
\end{equation}

\noindent where

\begin{equation}
  \alpha=\begin{cases}
  1.3 & 0.08M_{\odot} \leq M < 0.5 M_{\odot} \\
  2.3 & 0.5 M_{\odot} \leq M.
  \end{cases}
\end{equation}

\noindent As in our previous work we consider metallicities $Z$ and galactocentric 
distances $R_g$ of
$Z=0.005/R_g=2\,\rm{kpc}$, $Z=0.001/R_g=8\,\rm{kpc}$, and $Z=0.0002/R_g=20\,\rm{kpc}$, following the correlation of metallicities with distance observed for GCs in the Milky Way.  
Finally, we select 10\% of
particles to be binaries with separations consistent with {\"O}pek's law (flat
in $\log(a)$ from the point of stellar contact to the hard--soft boundary, where
the orbital velocity equals the typical velocity of particles in the cluster)
and thermal eccentricities ($p(e)de = 2e\,de$).  The primary masses are drawn
from the IMF, while the secondary masses are drawn from a uniform
distribution, $m_2 \in U[0,\,m_1]$.  

Our stellar evolution prescriptions are almost identical to those adopted in
\cite{Rodriguez2016a}.  We have added new physics describing mass loss from
pulsational-pair instabilities and pair-instability supernovae.  Briefly, we
follow \cite{Belczynski2016a}, and assume that any star whose pre-explosion
helium core mass is between $45\,M_{\odot}$ and $65\,M_{\odot}$ will undergo
pulsations that eject a significant amount of mass, until the
final product is at most $45 M_{\odot}$.  This yields a BH with a mass of $40.5\,M_{\odot}$, 
assuming that 10\% of the baryonic mass is lost during the conversion from baryonic to gravitational mass.
Stars with He core masses above $65 M_{\odot}$ are completely
destroyed in pair-instability supernovae.  There are significant uncertainties
on these numbers, and $40\,M_{\odot}$ is likely a conservative limit (with
many studies suggesting $\sim 50\,M_{\odot}$ for the lower limit of the mass gap
\cite[e.g.,][]{Woosley2017,Spera2017}).  However, our results here are
insensitive to the specific limit adopted for stellar-mass BH remnants. BHs in GCs will 
always
undergo mass segregation and form binaries from the most massive BHs first,
regardless of how large their masses are.  Furthermore, the GW recoil kicks applied to
merging BBHs are independent of the total mass (depending only on the mass ratio), 
implying that the retention of merger products also does not depend on the total mass.

{
One caveat must be addressed here.  By selecting $40.5M_{\odot}$ as our maximum BH mass, we are 
create a population of mergers with exactly equal-mass components 
($40.5M_{\odot}+40.5M_{\odot}$).  This may artificially increase the retention of 
BBH merger products, as equal-mass, non-spinning BBHs receive zero 
kicks.  We find that 40 of the 1521 in-cluster mergers are $40.5M_{\odot}+40.5M_{\odot}$ mergers. However, because these BBHs are the most massive in the cluster, they 
form and merge very early in the cluster lifetime, with the latest merger in any 
model occurring $\sim 1\rm{ Gyr}$ after cluster formation (at 
$z\sim2.3$).  GCs were significantly more massive and compact in the early 
universe, with correspondingly deeper central potentials and escape velocities.  
The lowest escape speed where a $40.5M_{\odot}+40.5M_{\odot}$ BBH merger occurred in our models was $31~
\rm{km}/\rm{s}$ (while $\sim 90\%$ of the mergers occurred where $v_{\rm{esc}} > 50~\rm{km}/\rm{s}$).  For non-spinning BHs and our adopted kick prescription, this 
would ensure the retention of all $40.5M_{\odot}+40.5M_{\odot}$ 
merger products assuming the true mass ratio was greater than $\sim0.9$}.

\end{document}